\documentclass[twocolumn]{jpsj3}

\usepackage{amsmath,amssymb}
\usepackage{graphicx}
\usepackage{xcolor}
\usepackage{color}

\begin{document}

\title{\textbf{Magnetic Polarization and Fermi Surface Instability: Case of YbRh$_2$Si$_2$}}
 

\author{
Alexandre~{\sc Pourret}$^1$\thanks{E-mail address: alexandre.pourret@cea.fr}, 
Georg~{\sc Knebel}$^1$\thanks{E-mail address: georg.knebel@cea.fr},
Tatsuma D.~{\sc Matsuda}$^2$, 
Gerard~{\sc Lapertot}$^1$,
and
Jacques~{\sc Flouquet}$^1$
}

\inst{%
$^1$SPSMS, UMR-E CEA / UJF-Grenoble 1, INAC, Grenoble, F-38054, France\\
$^2$Advanced Science Research Center, Japan Atomic Energy Agency, Tokai, Ibaraki 319-1195, Japan
}

\date{\today }

\abst{We report on thermoelectric and resitivity measurements in the antiferromagnetic heavy fermion compound YRh$_{2}$Si$_{2}$ at low temperatures and under high magnetic field. At low temperature, the thermoelectric power and the resistivity present several distinct anomalies as a function of field $H \parallel [110]$. A first anomaly appears at 3.5~T and a cascade of transitions above around $H_0 \sim 9.5$~T when the magnetic polarization reaches a critical value.  The anomalies at $H_0$ are accompanied with a change of sign of the thermoelectric power from negative  at low magnetic field ($H<H_0$) to positive at high field ($H>H_0$) and are resulting from a Lifshitz-type topological transition of the Fermi surface.   This field induced transition will be compared to the well characterized Fermi surface change in CeRu$_2$Si$_2$ at its pseudo-metamagnetic transition.
}


\maketitle

Heavy fermion compounds are recognized as key materials to study magnetic or valence quantum criticality \cite{Loehneysen2007, Flouquet2005}. Their low characteristic energy scale can be easily tuned by pressure or magnetic field. However, the electronic structure of these compounds is rather complicated and generally several bands are formed at the Fermi level. The flatness of the renormalized bands near the Fermi energy driven by the local character of the 4$f$ electrons adds the possibility of Fermi surface reconstructions under magnetic field either by crossing  the magnetic quantum criticality or entering into the high magnetic field Zeeman regime with a decoupling of majority and minority spin bands \cite{Daou2006}. 

Controversy still exists on the feedback between magnetic quantum criticality and Fermi surface instabilities. Basic questions on the theoretical modeling of quantum criticality are differently discussed. For example, in the spin-fluctuation approach, at first approximation the Fermi surface is assumed invariant on crossing an antiferromagnetic quantum critical point. The CeRu$_2$Si$_2$ series is considered as an excellent example for this model \cite{Flouquet2005, Knafo2009}. In difference in the local criticality approach,   Kondo break-down \cite{Coleman2001, Si2001, Pepin2007} has been involved with a drastic change of the Fermi surface from a small Fermi surface in the magnetically ordered state to a large Fermi surface in the paramagnetic heavy fermion state \cite{Paschen2004}. A archetype example of this scenario may be the heavy fermion compound YbRh$_2$Si$_2$ which orders antiferromagnetically below $T_{\rm N} = 0.07$~K \cite{Gegenwart2008}. This antiferromagnetic order is suppressed by a tiny magnetic field of $H_{\rm c} = 0.066$~T in the basal plane of the tetragonal crystal and a Fermi liquid behavior is restored above $H_{\rm c}$. Experimentally, the indications of an abrupt Fermi surface reconstruction at $H_{\rm c}$ are ambiguous as the huge value of the effective mass and the relative weakness of the critical field prevent direct quantum oscillations experiments through $H_{\rm c}$.  Only macroscopic quantities like the Hall effect \cite{Paschen2004, Friedemann2009} give some evidences for the reconstruction. However, this allows not  a straightforward determination of Fermi surface properties in this multiband system. The calculated Fermi surfaces of YbRh$_2$Si$_2$ consist of three main sheets, a hole-like surface centered at the Z point which has strong $f$ character, a second rather complex, multiconnected Fermi surface having both, electron and hole like contributions, and a small $\Gamma$ centered electron pocket \cite{Knebel2006, Wigger2007, Rourke2008, Friedemann2010, Zwicknagl2011}. 

In the present study we focus on the magnetic-field induced anomalies at around $H_0 \approx 9.5$~T applied along [110], i.~e.~much higher fields than $H_{\rm c}$. Above $H_0$ the specific heat coefficient $\gamma$ and the $A$ coefficient of the resistivity decreases significantly and the magnetization reaches an almost constant value near 0.9 $\mu_B$/mole Yb \cite{Tokiwa2005, Gegenwart2006, Knebel2006}. Fermi surface measurements by de Haas van Alphen (dHvA) technique have shown a strong variation of the observed frequencies originating from the hole like Fermi surface through $H_0 \approx 9.5$~T \cite{Rourke2008, Sutton2010}. 
While first a suppression of the local Kondo effect and a de-renormalisation of the quasiparticles occuring through $H_0$ have been proposed \cite{Gegenwart2006}, the changes of the dHvA frequencies have been interpreted in terms of a Lifshitz transition of the spin-splitted majority spin branch \cite{Rourke2008, Sutton2010}. A  recent band structure calculation using the renormalized band method concludes that the transition through $H_0$ results from a van-Hove-type structure in the density of states which moves with field through the Fermi level and drives a redistribution of the quasiparticles between different Zeeman-split Fermi surface sheets \cite{Friedemann2010, Zwicknagl2011}. 

In this article we report on detailed thermoelectric power (TEP) and magnetoresistivity measurements in YbRh$_2$Si$_2$ under magnetic field through $H_0$. As will be shown, several distinct anomalies in the TEP appear under magnetic field which originate from Fermi surface modifications.   
The high sensitivity of TEP to detect a topological electronical transitions is well established \cite{Varlamov1989}. The TEP is linked to the logarithmic derivative of the electronic conductivity $\sigma (\epsilon)$ at the Fermi level $\epsilon_{\rm F}$  
\begin{equation}
S=-\frac{\pi^2k_{\rm B}^2T}{3e}\left(\frac{\partial \ln \sigma (\epsilon)}{\partial \epsilon}\right)_{\varepsilon = \epsilon_{\rm F}}
\end{equation}
It contains two terms related to transport and thermodynamic properties of the conduction electron \cite{Barnard1972} 
\begin{equation}
S=-\frac{\pi^2k_{\rm B}^2T}{3e}\left(\frac{\partial \ln \tau (\epsilon)}{\partial \epsilon} +\frac{\partial \ln N(\epsilon)}{\partial \epsilon} \right),
\end{equation}
$\tau (\epsilon_{\rm F})$ and $N (\epsilon_{\rm F})$ being the relaxation time and the DOS of the quasiparticles at the Fermi level, respectively. Enhancement occurs for both terms proportional to $-\partial \ln N(\epsilon)/\partial \epsilon$ \cite{Miyake2005}. This derivative dependence differs  from the specific heat response which is proportional to $N(\epsilon)$ at low temperature. It may even lead to a change of sign of the TEP as its sign does not depend on that of the velocity but only on the curvature of the quasiparticle band. Despite this rather ''complex" response, it has been observed that at very low temperature the TEP is commonly positive for Ce heavy fermion compounds in agreement with the Kondo picture while it is negative for Yb heavy fermion compounds \cite{Behnia2004}. Furthermore, at $T \to 0$~K, the ratio of the TEP by the Sommerfeld coefficient $\gamma$ in SI units defines a dimensionless quantity $q$ inversely proportional to the heat carrier number \cite{Behnia2004} $q = \frac{S}{T}\frac{N_{\rm av}e}{\gamma}$ where $N_{\rm av}$ is the Avogadro number. 
An additional point is that heavy fermion compounds are multiband systems and the TEP power is the sum of the contributions of each subband weighted by their relative conductivity $S = \sum_{i}\frac{\sigma_i}{\sigma}S_i$ \cite{Miyake2005}.

The observed field dependence of the TEP in YbRh$_2$Si$_2$ will be compared to that of the well characterized case of CeRu$_2$Si$_2$ at its pseudo-metamagnetic transition at $H_{\rm m} = 7.8$~T. At $H_{\rm m}$ a drastic reconstruction of the Fermi surface occurs,\cite{Takashita1996, Aoki1993}  which is now characterized as a Lifshitz-type anomaly with a continuous shrinking of one spin-split majority Fermi surface until it is driven below the Fermi surface at $H_{\rm m}$  \cite{Daou2006}. This transition is accompanied with sharp extrema in the field dependence of the TEP \cite{Amato1989, Pfau2012, Machida2013}.

\begin{figure}[ct!]
\begin{center}
\includegraphics[width=7.3cm]{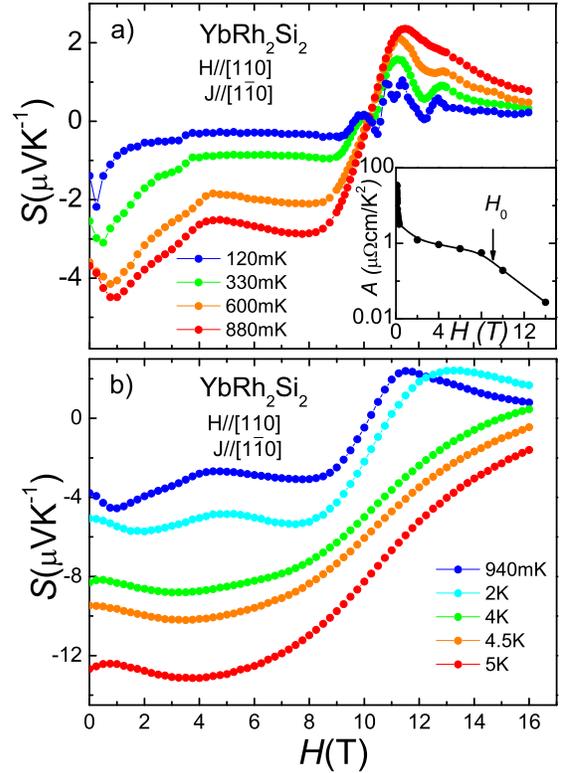}	
\end{center}
\caption{(Color online) Thermoelectric power $S$ of YRh$_{2}$Si$_{2}$ for $J\!\!\parallel \!\! [1\bar{1}0]$ and $H \!\!\parallel \!\![110]$ as a function of magnetic field for temperatures (a) below and (b) above 900mK. At very low temperature, at least seven extrema can be resolved near $H_{0}$ corresponding to a change of sign of the TEP. With increasing temperature above 1~K the extrema get less pronounced and broaden. At high temperatures, $S$ stays negative on all the field range. The inset shows the field dependence of the $A$ coefficient of the resistivity \cite{Knebel2006}.}\label{fig1}
\end{figure}

TEP and magnetoresistivity experiments have been performed on the same whisker-like out of In flux grown single crystal of YbRh$_2$Si$_2$ with dimensions of 4.3~mm x 0.15~mm x 0.15 mm. The magnetic field is applied along [110] of the tetragonal crystal and heat or charge current perpendicular to the field along $[1\bar{1}0]$, which corresponds to the long axis of the crystal. The high quality of the single crystal is confirmed by the observation of Shubnikov de Haas oscillations above 10~T. TEP experiments have been performed by a one heater, two thermometers set-up in a dilution refrigerator down to 120~mK and up to 16~T. Thermometers and heater are thermally decoupled from the sample holders by highly resistive manganin wires. Resistivity measurements have been performed down to 30~mK and fields up to 13~T by a four point lock-in technique using a low temperature transformer. In this paper we do not focus on the field suppression of the antiferromagnetic order at $H_{\rm c}$. 

Figure \ref{fig1} (a) shows the field dependence of the TEP of YbRh$_2$Si$_2$ at various temperatures from 120~mK up to 880~mK. The field response of the TEP is rather complex.  Clearly several anomalies appear between 0 and 16~T. All signatures in the TEP gets sharper with lowering the temperature. At $H \approx 3.5$~T $S (H)$ increases steplike and around $H \approx 10$~T at least four sharp maxima appear. The step-like anomaly at $H \approx 3.5$~T is quite different from the extrema observed in the field range (9~T - 12~T) around $H_0$. It may be connected to a drastic change in the interplay between inter-site magnetic correlations and the on-site local Kondo fluctuations as shown in the inset of Fig.~\ref{fig1}~(a) which displays the  field dependence of $A(H)$. Next we focus on the anomalies around $H_0$. Remarkably,  the sign of the TEP changes from negative ($H < H_0$) to positive  ($H > H_0$).  With increasing temperature above 1~K the extrema get less pronounced and broaden (see Fig.~\ref{fig1}~(b)). As indicated in the inset of Fig.~\ref{fig1}~(a) the $A$ coefficent of the resistivity ($\sqrt{A} \propto \gamma$, the Sommerfeld coefficient of the specific heat) has a sharp maximum at $H_{\rm c} \approx 70$~mT and decreases continuously as the field increases with a drastic drop through $H_0$. 

\begin{figure}[t!]
\begin{center}
\includegraphics[width=8cm]{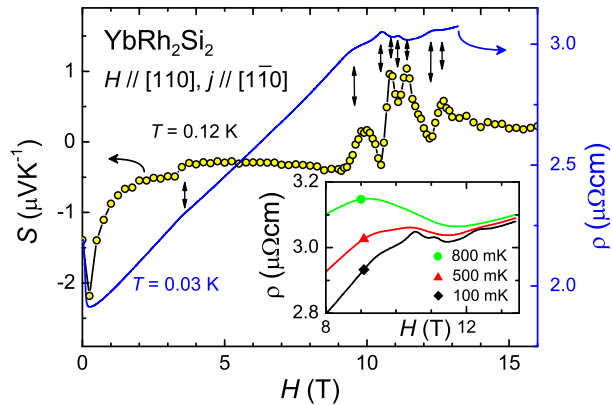}	
\end{center}
\caption{(Color online) Thermoelectric power $S$ and magnetoresistance $\rho (H)$ of YRh$_{2}$Si$_{2}$ for $H \!\!\parallel \!\![110]$ and $J\!\!\parallel \!\![1\bar{1}0]$ in the basal plane as a function of magnetic field measured at 120~mK and 30~mK respectively. All the different anomalies (extrema and inflection points) are indicated by double-vertical arrows underlying the excellent agreement between TEP and resistivity measurements. The inset shows the $\rho (H)$ at various temperatures in the field range around $H_0$ in an enlarged view. }\label{fig2}
\end{figure}

Another remarkable feature is presented in the magnetoresistivity $\rho (H)$ shown in Fig.~\ref{fig2}, which displays also several anomalies crossing through $H_0$. The fields of the anomalies (extrema and inflection points) in $\rho (H)$ coincide perfectly with the sharp anomalies in the TEP. The inflection point of the drop of $\rho (H)$ at low fields defines the highly discussed crossover scale $T^\star$ \cite{Gegenwart2008, Friedemann2009}. At $H \sim 3.5$~T the step in the TEP corresponds to a kink in $\rho (H)$. The inset in Fig.~\ref{fig2} displays $\rho (H)$ for different temperatures around $H_0$ in an enlarged view and the smearing of the anomalies with increasing temperature is clearly demonstrated. This indicates that the anomalies are driven by Fermi surface instabilities. 

\begin{figure}[t!]
\begin{center}
\includegraphics[width=7.3cm]{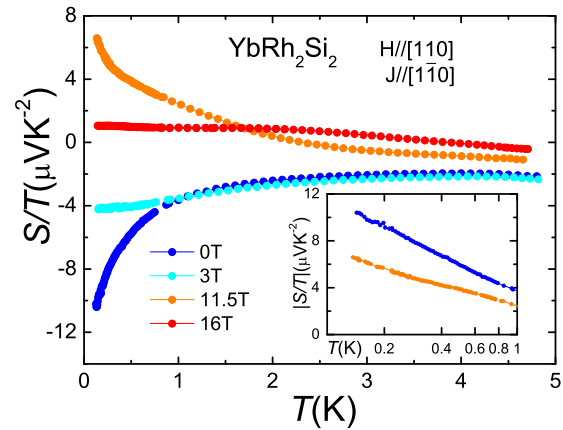}	
\end{center}
\caption{(Color online) Thermoelectric power divided by the temperature $S/T$ of YRh$_{2}$Si$_{2}$ for $J\!\!\parallel \!\![1\bar{1}0]$ and $H \!\!\parallel \!\![110]$ as a function of temperature for different magnetic fields. The inset shows that $S/T$ for $H=0$~T and 11.5~T show a logarithmic temperature dependence over more than one decade.}\label{fig3}
\end{figure}

Figure \ref{fig3} shows the temperature dependence of the TEP divided by the temperature $S/T$ at various fields on a broad temperature range from 120~mK up to 5~K. At zero field the TEP shows clearly a logarithmic increase on cooling down to the lowest temperature of our experiment of 120~mK (see inset of Fig.~\ref{fig3}) due to the non Fermi liquid behavior in agreement with the previous reports \cite{Hartmann2010, Machida2012}.  However, below 100~mK  $S/T$ has been observed to drop down abruptly in the low field regime $H \leqslant H_c$, while it gets constant only in the Fermi liquid regime $H > H_c$\cite{Hartmann2010, Machida2012, Kim2011}.  Enhanced values of $S/T$ have been reported for a few heavy-fermion systems and are considered to be in close relation to the strongly enhanced specific heat in these systems on approaching the magnetic quantum criticality\cite{Behnia2004}. When approaching the critical field $H_{c}$ of YbRh$_{2}$Si$_{2}$ from the high field side ($H>H_c$) the $A$-coefficient of the $T^{2}$ term to $\rho(T)$ (inset Fig.~\ref{fig1}) and the $\gamma$ value are strongly increasing. Interestingly the temperature dependence of the TEP in the intermediate temperature regime of 120~mK to 1~K taken at the maximum of $S$ at $H = 11.5$~T (just above $H_{0}$) is quite comparable, despite the opposite sign of the TEP at low temperature, to that measured at $H=0$ in the same $T$ regime. However, here the $-\log T$ behavior is not a precursor of the proximity of magnetic quantum criticality but linked to the Zeeman instability of the renormalized bands.

\begin{figure}[ct!]
\begin{center}
\includegraphics[width=8cm]{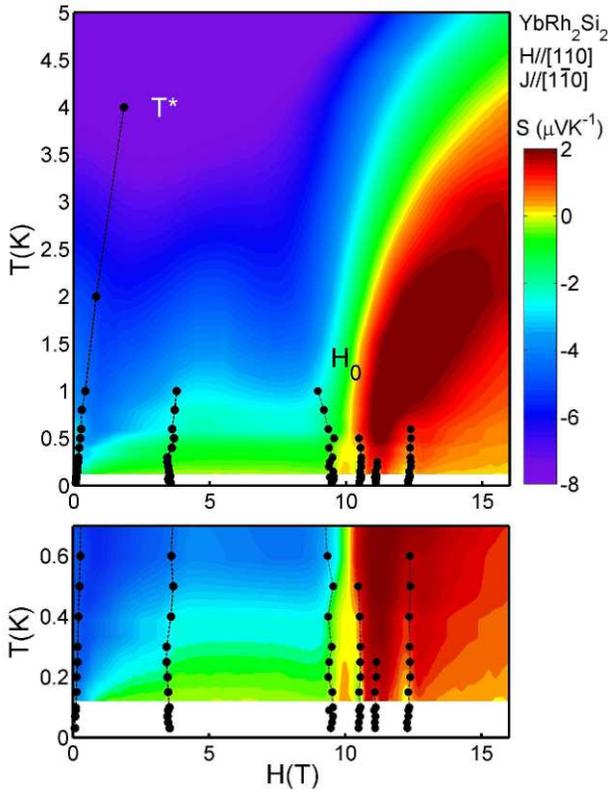}	
\end{center}
\caption{(Color online) Linear color map of the thermoelectric power in the ($H,T$) plane for temperature between 120~mK and 5~K (a) and between 120~mK and 900~mK (b). Superimposed on the plot are the values of the different anomalies in resistivity (black circles). The low field anomaly correspond to $T^\star$, the position of the critical field $H_{0}$ corresponding to a change of sign in the TEP at low temperature and a maximum in the resistivity is emphasized (dashed line). }\label{fig4}
\end{figure}

The results of the TEP measurements are summarized in the color plot Fig.~\ref{fig4} which shows $S (T,H)$ in an $(H,T)$ phase diagram. In addition the positions of the critical fields of the magnetoreistance are indicated. $S$ is negative at low fields as expected for a Yb based heavy fermion compound. On approaching $H_0$ the sign of the TEP changes from negative to positive. As indicated in Fig.~\ref{fig4} (b) the position $H_0$ is obviously well defined at low temperatures and will move only at temperatures above 900~mK. 

In agreement with dHvA experiments \cite{Rourke2008, Sutton2010} it is clear that the topological electronic instability concerns several subbands as marked by the cascade of maxima in the field range from 9.5~T to 12~T.  In previous dHvA experiments several orbits from the $Z$ centered hole Fermi surface which has predominant $f$ character have been observed, while only one orbit of the ''J" sheet has been detected above 13~T. It is also reasonable to believe that in TEP the major signal above $H_0$ will come from this hole Fermi surface while the negative sign of the TEP below $H_0$ indicates electronic main charge carriers. This abrupt change of the main heat carriers is a clear signature of the topological change of the Fermi surface.  

\begin{figure}[ct!]
\begin{center}
\includegraphics[width=8cm]{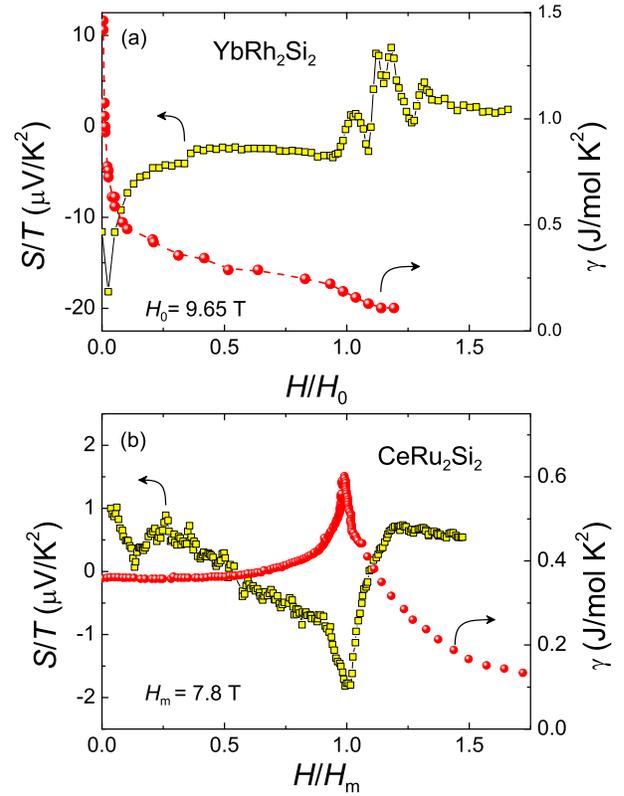}	
\end{center}
\caption{(Color online) (a) Field dependence ($H\parallel [1\bar{1}0]$) of $S/T$ and specific heat coefficient $\gamma$ (taken from Ref.\citen{Gegenwart2006}) (right axis) of YbRh$_2$Si$_2$ as function of field $H$ the normalized to $H_0 = 9.65$~T (left axis). (b) Field dependence ($H\parallel c$) of $S/T$ (taken from ref.~\citen{Pfau2012}) and $\gamma$ (see ref.~\citen{Flouquet2010}) of CeRu$_2$Si$_2$  normalized to the metamagnetic transition at $H_{\rm m} = 7.8$~T. Let us remark that the size of the Lifshitz anomaly between the two cases scales roughly with their differences in the initial $\gamma$ values. }\label{fig5}
\end{figure}

It is worthwhile to compare the field variation of the TEP of YbRh$_2$Si$_2$ through $H_0$ with the results obtained for CeRu$_2$Si$_2$ at the pseudo-metamagnetic transition $H_{\rm m}$. Figure~\ref{fig5} shows the respective field dependence of $S/T$ and $\gamma$ for both compounds. In both cases a distinct modification of the Fermi surface occurs, when the magnetic polarization of the 4$f$ electrons reaches a critical value. The simplest image is that the cross section of the majority spin band increases leading to an instability on approaching the Brillouin zone and thus drive the electronic transition. Both systems, CeRu$_2$Si$_2$ and YbRh$_2$Si$_2$, are compensated metals with rather large Fermi surfaces. In CeRu$_2$Si$_2$ only one extremum of $S/T$ is detected. Here, the pseudo-metamagnetic transition is associated to  huge magnetostriction effect ($\Delta V/V \sim 10^{-3}$) while in YbRh$_2$Si$_2$ the feedback with the volume change is far weaker ($\sim 10^{-5}$) and occurs over a large $H$ region. Furthermore as in CeRu$_2$Si$_2$ the pseudo-metamagnetic field $H_{\rm m} \sim 7.8$~T is near the value of the metamagnetic critical end field ($H_{\rm c}^\star \sim 4$~T) \cite{Flouquet2010}, antiferromagnetic, ferromagnetic and local interactions are strongly interfering at $H_{\rm m}$ leading to a huge enhancement of the $\gamma$ value at $H_{\rm m}$. In difference, in YbRh$_2$Si$_2$, $H_0$ is strongly decoupled from the critical field $H_{\rm c}$ of the collapse of the antiferromagnetic order, thus it is possible to separate different instabilities, the antiferromagnetic and the Zeeman driven ones. 

As in YbRh$_2$Si$_2$ the ratio $H_0/H_{\rm c} \sim 10^2$, the main mechanism which gives the mass renormalization is the Kondo lattice effect as shown by the proportionality of the energy scales of the Kondo temperature ($T_{\rm K} \sim 25$~K)  and $H_0$. The collapse of the magnetic interactions and thus the suppression of the magnetic quantum criticality under field leads to a $\gamma$ term of roughly 300~mJ/mol K$^2$ before droping at $H_0$. Extrapolating to $H=0$ by neglecting the magnetic correlations suggests a Sommerfeld coefficient near 670~mJ/mol K$^2$ \cite{Zwicknagl2011}. Let us remark that (i) the mass enhancement at the quantum critical point $\gamma (H_{\rm c})/\gamma (H = 0)$ never exceeds a strong factor (2 or 3), (ii) the equivalence of the Kondo and Zeeman scales $k_{\rm B} T_{\rm K} \sim g\mu_{\rm B} H_0$ suggests that the Zeeman decoupling will appear at a critical value of the magnetization by the Kondo effect, assuming the susceptibility $\chi \propto 1/T_{\rm K}$ and thus a constant magnetization $M_0 \propto H_0/T_{\rm K} \sim$~const. Experimentally, an almost constant magnetization $M_0 \sim 0.9 \mu_{\rm B}$ in YbRh$_2$Si$_2$ is achieved above $H_0$ \cite{Gegenwart2006}. At least  there is clearly a critical size of the magnetic polarization where a Fermi surface instability occurs and this may drive the huge decrease of the mass renormalization and $\gamma$ above $H_0$.

Finally we want to underline the interest of macroscopic probes like TEP as they are able to detect fine structures in continuous field scans while quantum oscillations experiments require a finite field window to detect by Fourier transformation a change in frequency. 

Recent theoretical developments on the role of the Zeeman energy to drive a Lifshitz transition have been reported \cite{Kusminskiy2008, Hackl2011}. A specific study on the YbRh$_2$Si$_2$ case has been realized by looking to the magnetic field induced changes of the heavy quasiparticle described in realistic renormalized band model \cite{Zwicknagl2011}. As observed here, the anomalies at $H_0 \sim 9.5$~T result from a van Hove singularity in the DOS which appears below the Fermi level at $H = 0$ which can be revealed in magnetic field. It is this combined effect of local many body effects and coherence given by the periodicity of the lattice which produces this van Hove singularity \cite{Zwicknagl2011}. However the present theoretical work in magnetic field is restricted to the analysis of the quasiparticle density of states, a further step will be to indicated the Fermi surface change.\cite{Pfau2013a}

In conclusion we showed that TEP is an excellent probe to detect topological transitions of the Fermi surface. We observed in YbRh$_2$Si$_2$ several sharp anomalies in the field dependence of the TEP which has been interpreted as topological Lifshitz transitions. This transitions appear when the magnetization reaches a critical value as it has been previously established for CeRu$_2$Si$_2$. 

\acknowledgement{This work has been supported by the French ANR project PRINCESS and the ERC starting grant NewHeavyFermion.}




\end{document}